\documentclass{article}

\usepackage{arxiv}
\makeatletter
\renewcommand\section{\@startsection{section}{1}{\z@}%
  {-2.5ex \@plus -1ex \@minus -.2ex}
  {0.8ex \@plus .2ex}
  {\normalfont\Large\bfseries}}
\makeatother

\usepackage[utf8]{inputenc} 
\usepackage[T1]{fontenc}    

\usepackage{hyperref}       
\usepackage{url}            
\usepackage{booktabs}       
\usepackage{amsfonts}       
\usepackage{amsmath}
\usepackage{amssymb}
\usepackage{microtype}      
\usepackage{graphicx}
\usepackage{lipsum}
\usepackage[numbers]{natbib} 
\usepackage{doi}

\hypersetup{
  colorlinks = true,
  linkcolor  = black,  
  citecolor  = blue,   
  urlcolor   = blue    
}

\title{Seismic Wave Propagation in Viscoelastic Media under Atangana–Baleanu Fractional Dynamics: Model Formulation and Numerical Simulations}

\author{ \href{https://orcid.org/0000-0001-7402-4468}{\includegraphics[scale=0.06]{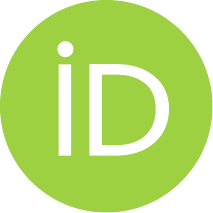}\hspace{1mm}Taylan Demir}\thanks{Use footnote for providing further
		information about author (webpage, alternative
		address)---\emph{not} for acknowledging funding agencies.} \\
	Department of Mathematics\\
	  Ankara University\\
	Ankara, Turkey \\
	\texttt{taylandemir@ankara.edu.tr} \\
	\And
	\href{https://orcid.org/0009-0005-6060-0547}{\includegraphics[scale=0.06]{orcid.pdf}\hspace{1mm}Atakan Koçyiğit} \\
	Department of Electrical and Electronics Engineering\\
	  Atılım University\\
	Ankara, Turkey \\
	\texttt{atakan.kocyigit@megetek.com.tr} \\
}

\renewcommand{\shorttitle}{\textit{arXiv} Template}

\hypersetup{
pdftitle={A template for the arxiv style},
pdfsubject={q-bio.NC, q-bio.QM},
pdfauthor={David S.~Hippocampus, Elias D.~Striatum},
pdfkeywords={First keyword, Second keyword, More},
}
\DeclareUnicodeCharacter{2212}{-}

\begin{document}
\maketitle

\begin{abstract}
We propose a one-dimensional viscoelastic seismic-wave model driven by the Atangana-Baleanu-Caputo fractional derivative with a non-singular Mittag-Leffler kernel. A finite-difference discretization in space and an Adams-Bashforth-Moulton predictor-corrector scheme in time are used to compute solutions for several fractional orders. Simulations indicate that fractional memory alters both attenuation and dispersion, leading to non-exponential energy decay compared with the classical integer-order case.
\end{abstract}
\section{Introduction}
Seismic waves travelling through realistic geological media are affected by the viscoelasticity of the medium, it also exhibits microstructure heterogeneity, and temporal memory, and thus the seismic signals that result will display an attenuated and dispersive characteristic that is different than that described by classical integer order models with an exponential relation for attenuation.
In addition in using classical first order time derivative relations to obtain stress–strain curves are unable to adequately account for the lasting nature of seismic energy in order to predict the nearly power-law decay that has been observed in laboratory and field data [2].
To provide a systematic treatment of memory-driven effects, fractional calculus has been applied extensively in Physics and Continuum Mechanics, with applications compiled in [1] and addressed in [2].
Replacing integer-order time derivatives in the constitutive equations with non-integer-order operators allows for the introduction of nonlocal kernels that encode a hereditary type of behaviour within the constitutive equations while providing a differential form consistent with the standard balance equations. In the area of viscoelasticity, many fractional viscoelastic models have been employed to predict the stress relaxation, creep and frequency-dependent attenuation associated with viscoelastic behaviour. Examples of these models include the linear viscous damping resistant model of Caputo [4], which has a nearly frequency-independent quality factor; and the modelling of the viscoelasticly damped structures investigated by Bagley and Torvik [5]. The Atangana--Baleanu (AB) derivative defined in the Caputo sense through [3] has proven to be an effective tool when modeling seismic activity, as the derivative is defined using a nonlocal Mittag--Leffler kernel that is non-singular at the origin. This allows for a smoother interpolation between the two extreme behaviors of known seismic operations (i.e., purely elastic vs. strong dissipation) than what is available using traditional Riemann--Liouville or Caputo Kernels. Therefore, our purpose here is to create a fractional viscoelastic model to explain how seismic waves propagate in one-dimensional space. This is accomplished by creating a fractional wave equation that allows the model to account for the short-time elastic response and long-term dissipative processes within the same mathematical expression. We have developed a numerical simulation framework based on an Adams-Bashforth-Moulton predictor-corrector approach with an adapted Adams-Bashforth kernel and applied a conventional mathematical treating methods in a spatial domain. In addition, we have applied the new mathematical treating methods to comparing both classical and fractional dynamic systems through displacement and energy decay with varying levels of fractional order to examine the approach's ability to replicate the non-exponential / nearly power-law attenuation of seismic waveforms resulting from the damping characteristics of viscoelastic materials. The discrete wavelet transform (DWT) and the Atangana-Baleanu fractional derivative represent two different mathematical tools that two different approaches can utilize to analyze engineering systems. The DWT offers localized analysis of time-frequency data, thus being a powerful tool to extract the features of time-series data; whereas, the AB derivative incorporates past states' influence on a system's current dynamic. Therefore, DWT is primarily used to analyze data while the AB derivative models are primarily used for simulation or models, either ML or physics based. While DWT and AB derivative can both be implemented with MATLAB or Python, but due to engineering considerations, Python offers more significant advantages. Thanks to the extensive library systems available along with a very high degree of computational power associated with Python's open-source structure, researchers can use Python to effectively manipulate and analyze large datasets. In addition to offering many benefits due to its open-source nature and the availability of software libraries that provide support for the analysis of large datasets (e.g., Pandas and NumPy), Python's primary advantage over other programming languages, such as MATLAB, lies in his/her ability to manipulate data rapidly and efficiently than that of MATLAB. Also, because Python is open-source and therefore does not require a license to use, researchers will not be limited by licensing restrictions, can easily share their code with other researchers, and increase the reproducibility of their research results; furthermore, this capability is especially important when utilizing new mathematical operations, such as Atangana–Baleanu derivatives, since it facilitates rapid prototype development and encourages collaboration. Python’s multi-faceted programming paradigm of OOP, functional, and procedural can create an Object-AB operator using classes and provides a straightforward and uncomplicated means for future extension to use various initial states and kernel functions. Due to its vast potential with regards to performance when applied to these projects (and also due to the ability to use multiple machines to conduct large-scale simulations), Python has gained popularity in data processing applications involving large data sets such as signal processing (DSP) and control algorithms and the AB operator. In conjunction with machine learning, the advantages of the Atangana-Baleanu derivative can produce even greater effects when memory effects are modeled for physical systems. Machine learning thus provides a unique method of validating solutions for highly complex and nonlinear differential equations and accelerating their development. A considerable body of literature clearly establishes the ability of these techniques to accurately model and verify very complicated nonlinear systems [6]. Therefore, taking into consideration all these factors, Python provides the ideal balance between flexibility, computational efficiency, and being open-source compared to other popular programming languages for engineering applications.

\section{Preliminaries and Classical Model}
The discrete wavelet transform (DWT) and the Atangana-Baleanu (AB) fractional derivative are complementary tools used in the analysis and modeling of engineering systems that have a high degree of complexity. The DWT gives a time-frequency representation of the signal, thus it is useful for multiresolution analysis, denoising, and extracting features from signals or images that are subject to noise [7]. AB Fractional Operators allow for modelling of Memory Effects (including Hereditary) through use of a Nonlocal/Nonsingular Mittag-Leffler Kernel. This approach provides another means of representing memory effects than via Classical Fractional Calculus (using a Power Law Kernel) [1,2,3]. Therefore, from the Modeling Point of View, the DWT is better suited for diagnosing and performing Feature Engineering using empirical or data-driven approaches. However, when you are implementing something that needs to take into account both Hereditary Effects and the Anomalous Dissipation of Energy, it makes sense to use the AB Fractional Derivative to create simulations based on Models that account for these types of dynamics [3,4,5]. Both Modeling Techniques can be implemented efficiently in MATLAB and Python. While MATLAB has existing, commercially available, established toolboxes available for both Wavelet Analysis and Numerical Solutions of Differential Equations, Python provides a wide range of flexibility and an open-source approach via its ecosystem of open-source libraries (such as NumPy, SciPy and PyWavelets) that support large scale simulated modelling and easy integration with the latest scientific computer software workflows. This greater flexibility supports rapid prototyping of ideas using newly available fractional calculus operators (like the AB-derivative), as well as allowing users to quickly share their concepts through reproducibly generated Jupyter notebooks, and to connect these notebooks with symbolic mathematics and/or data processing libraries [1,2]. Specifically, DWT-based preprocessing (i.e. decomposing) may effectively denoise/compress experimental signal data, while the use of AB-type fractional processes may ¿t capture the inherent viscoelastic or diffusion processes of the underlying dynamic system, which allows for smooth flow from the original experimental signal to calibrated fractional models. Recent research shows that Atangana-Baleanu fractional derivatives, a type of non-singular fractional operator, can effectively model the behaviour of complex nonlinear and chaotic systems. For example, fractal-fractional versions of the Atangana-Baleanu operator have been utilized to design and analyze chaotic electronic circuits, allowing for accurate representation of hidden oscillations in addition to providing ways to model multi-stable states and complicated attractors [6]. These findings provide further support for previous efforts that applied fractional derivatives with singular kernels to develop models of viscoelastic damping and essentially frequency independent dissipation within solid materials and structures [1,4,5]. All of these findings provide strong motivation for employing the Atangana-Baleanu fractional derivative in conjunction with wavelet methods, as a broad-based model for analyzing and simulating complex and multiscale engineered systems that exhibit significant memory properties.
\section{Atangana--Baleanu fractional viscoelastic model}

In classical one–dimensional linear viscoelasticity, wave propagation in a bar of density $\rho$ is governed by the balance of linear momentum
\begin{equation}
  \rho\,\frac{\partial^{2}u}{\partial t^{2}}(x,t) = \frac{\partial\sigma}{\partial x}(x,t),
  \qquad 0<x<L,\ t>0,
\end{equation}
where $u(x,t)$ denotes the longitudinal displacement and $\sigma(x,t)$ the Cauchy stress. The small strain is given by $\varepsilon(x,t)=\partial u/\partial x$. For materials exhibiting power–law type attenuation and dispersion, it is natural to introduce fractional derivatives into the constitutive relation between $\sigma$ and $\varepsilon$, leading to fractional diffusion–wave equations that generalize classical Kelvin–Voigt or Zener models [4, 5, 8].

To account for a non–singular memory effect, we adopt the Atangana--Baleanu fractional derivative in the Caputo sense. For a sufficiently smooth function $f:[0,T]\to\mathbb{R}$ and order $0<\alpha<1$, the Atangana--Baleanu–Caputo (ABC) derivative is defined by
\begin{equation}
  ({}^{ABC}D_{t}^{\alpha} f)(t)
  = \frac{B(\alpha)}{1-\alpha}\int_{0}^{t}
      E_{\alpha}\!\left(-\frac{\alpha}{1-\alpha}(t-\tau)^{\alpha}\right)
      f'(\tau)\,d\tau,
  \qquad 0<t\le T,
\end{equation}
where $E_{\alpha}$ is the one–parameter Mittag–Leffler function and $B(\alpha)$ is a normalization function satisfying $B(0)=B(1)=1$ [3]. The Mittag–Leffler kernel is non–singular at $t=0$ and decays more slowly than an exponential, which allows the operator to interpolate between almost elastic short–time response and long–time diffusive–like relaxation in diffusion and heat–transfer models described by the same derivative [3, 9].

In analogy with classical fractional viscoelasticity, we model the constitutive law of the viscoelastic medium by superposing an instantaneous elastic response with modulus $E_{0}>0$ and a hereditary contribution driven by the ABC derivative. Denoting the second modulus by $E_{1}\ge0$, we write the linear stress–strain relation in the form
\begin{equation}
  \sigma(x,t)
  = E_{0}\,\varepsilon(x,t)
    + E_{1}\,({}^{ABC}D_{t}^{\alpha}\varepsilon)(x,t),
  \qquad \varepsilon(x,t)=\frac{\partial u}{\partial x}(x,t).
\end{equation}
Such linear laws, with different choices of fractional operators, are standard in the modeling of viscoelastic damping and reproduce experimentally observed power–law rheology [4, 5, 8], while the choice of the Atangana--Baleanu kernel provides a non–singular alternative to classical power–law kernels [3, 9].

Substituting the above constitutive relation into the momentum balance and using $\varepsilon=u_{x}$, we obtain
\begin{equation}
  \rho\,\frac{\partial^{2}u}{\partial t^{2}}(x,t)
   = \frac{\partial}{\partial x}
     \Big(
       E_{0}\,\frac{\partial u}{\partial x}(x,t)
       + E_{1}\,{}^{ABC}D_{t}^{\alpha}\!\Big[\frac{\partial u}{\partial x}(x,t)\Big]
     \Big).
\end{equation}
Assuming $E_{0}$ and $E_{1}$ are constant and interchanging the order of the spatial derivative with the ABC operator, this yields the fractional viscoelastic wave equation
\begin{equation}
  \rho\,\frac{\partial^{2}u}{\partial t^{2}}(x,t)
  = E_{0}\,\frac{\partial^{2}u}{\partial x^{2}}(x,t)
    + E_{1}\,{}^{ABC}D_{t}^{\alpha}
        \Big[\frac{\partial^{2}u}{\partial x^{2}}(x,t)\Big].
\end{equation}
Introducing the characteristic wave speed $c_{0}^{2}=E_{0}/\rho$ and the dimensionless parameter $\kappa=E_{1}/E_{0}$, we can rewrite the equation in the normalized form
\begin{equation}
  \frac{\partial^{2}u}{\partial t^{2}}(x,t)
  - c_{0}^{2}\,\frac{\partial^{2}u}{\partial x^{2}}(x,t)
  - \kappa\,c_{0}^{2}\,
    {}^{ABC}D_{t}^{\alpha}\!\Big[\frac{\partial^{2}u}{\partial x^{2}}(x,t)\Big]
  = 0,
  \label{eq:AB_wave}
\end{equation}
which belongs to the general class of wave equations with fractional viscoelastic damping studied in the recent literature [10, 11].

The model is supplemented with initial conditions
\begin{equation}
  u(x,0)=u_{0}(x),\qquad
  \frac{\partial u}{\partial t}(x,0)=v_{0}(x),
\end{equation}
and with physically appropriate boundary conditions, for instance fixed ends
$u(0,t)=u(L,t)=0$ or traction–free boundaries $\sigma(0,t)=\sigma(L,t)=0$. By introducing the dimensionless variables
\[
  \xi = \frac{x}{L},\qquad \tau = \frac{c_{0} t}{L},
\]
and rescaling the displacement as $u(x,t)=U_{0}w(\xi,\tau)$, equation \eqref{eq:AB_wave} transforms into
\begin{equation}
  \frac{\partial^{2}w}{\partial \tau^{2}}(\xi,\tau)
  - \frac{\partial^{2}w}{\partial \xi^{2}}(\xi,\tau)
  - \kappa\,{}^{ABC}D_{\tau}^{\alpha}
      \Big[\frac{\partial^{2}w}{\partial \xi^{2}}(\xi,\tau)\Big]
  = 0,
\end{equation}
posed on $0<\xi<1$, $\tau>0$. In this formulation the parameter $\kappa$ measures the relative strength of viscoelastic memory. For $\kappa=0$ the model reduces to the classical undamped wave equation, while increasing $\kappa$ and/or the order $\alpha$ enhances the long–time damping and dispersion induced by the non–singular Atangana--Baleanu kernel, in line with observations reported for diffusion and wave processes in complex viscoelastic media [3, 9, 10, 11].
\section{Numerical scheme: Adams--Bashforth--Moulton discretisation}

For the numerical approximation of the fractional viscoelastic wave equation we first apply a method-of-lines strategy. The spatial interval $(0,L)$ is partitioned into $N$ subintervals of length $h=L/N$ and we denote by $x_{i}=ih$, $i=0,\dots,N$, the grid points. Let $u_{i}(t)\approx u(x_{i},t)$ and approximate the second spatial derivative by the standard centred finite difference operator
\[
  (\delta_{xx} u)_{i}(t) = \frac{u_{i+1}(t)-2u_{i}(t)+u_{i-1}(t)}{h^{2}},
  \qquad i=1,\dots,N-1.
\]
Introducing the vectors $U(t)=(u_{1}(t),\dots,u_{N-1}(t))^{T}$ and $(\delta_{xx}U)(t) = ((\delta_{xx}u)_{1}(t),\dots,(\delta_{xx}u)_{N-1}(t))^{T}$, the dimensionless fractional wave equation derived in the previous section can be written in semi-discrete form as
\begin{equation}\label{eq:semi_discrete}
  \ddot U(t) = A U(t) + \kappa A\, ({}^{ABC}D_{t}^{\alpha} U)(t),
\end{equation}
where the dot denotes differentiation with respect to $t$, $A$ is the symmetric negative definite matrix associated with the second-order finite difference operator, and ${}^{ABC}D_{t}^{\alpha}$ is the Atangana--Baleanu--Caputo (ABC) derivative of order $0<\alpha<1$ acting componentwise on $U$ [4,9]. Equation \eqref{eq:semi_discrete} represents a system of second-order fractional ordinary differential equations with a non-singular memory kernel. In order to apply a one-step fractional predictor--corrector method, it is convenient to rewrite it as a first-order system by setting $V(t)=\dot U(t)$ and $Y(t)=(U(t),V(t))^{T}$; then the system can be cast in the abstract form
\begin{equation}\label{eq:FDE_abstract}
  \frac{dY}{dt}(t) = F\bigl(t,Y(t),{}^{ABC}D_{t}^{\alpha}Y(t)\bigr),
\end{equation}
where the right-hand side $F$ contains the elastic and viscoelastic contributions generated by $A$ and the ABC derivative.

The ABC derivative in the Caputo sense admits an integral representation with a non-singular Mittag--Leffler kernel [3]. For a sufficiently smooth vector-valued function $Z(t)$ this can be written as
\[
  ({}^{ABC}D_{t}^{\alpha}Z)(t)
  = \frac{B(\alpha)}{1-\alpha}\int_{0}^{t}
      E_{\alpha}\!\left(-\frac{\alpha}{1-\alpha}(t-\tau)^{\alpha}\right)
      Z'(\tau)\,d\tau,
\]
where $E_{\alpha}$ is the one-parameter Mittag--Leffler function and $B(\alpha)$ is a normalisation factor with $B(0)=B(1)=1$. This Volterra integral form allows us to adapt the classical Adams--Bashforth--Moulton (ABM) predictor--corrector method for Caputo-type equations, as developed in [2,12,13], to the present non-singular kernel. We introduce a uniform temporal grid $t_{n}=n\Delta t$ for $n=0,1,\dots,N_{t}$ with time step $\Delta t>0$, and denote $Y^{n}\approx Y(t_{n})$ and $F^{n}=F(t_{n},Y^{n},{}^{ABC}D_{t}^{\alpha}Y^{n})$. Following [12,13], the ABM scheme consists of two stages. In the explicit predictor step one computes a provisional value $Y^{n+1,P}$ by means of a fractional Adams--Bashforth quadrature,
\begin{equation}\label{eq:predictor}
  Y^{n+1,P}
  = Y^{0}
    + \Delta t^{\alpha}\sum_{j=0}^{n} b_{j}^{(n+1)} F^{j},
\end{equation}
where the weights $b_{j}^{(n+1)}$ depend on the fractional order $\alpha$ and are obtained by integrating the ABC kernel over the subintervals $[t_{j},t_{j+1}]$. In the implicit corrector step, an Adams--Moulton-type formula is applied to refine the approximation at $t_{n+1}$,
\begin{equation}\label{eq:corrector}
  Y^{n+1}
  = Y^{0}
    + \Delta t^{\alpha}
      \left(
        a_{0}^{(n+1)} F\bigl(t_{n+1},Y^{n+1,P},{}^{ABC}D_{t}^{\alpha}Y^{n+1,P}\bigr)
        + \sum_{j=0}^{n} a_{j+1}^{(n+1)} F^{j}
      \right),
\end{equation}
where the weights $a_{j}^{(n+1)}$ are the standard Adams--Moulton fractional quadrature coefficients adapted to the ABC kernel. Explicit expressions for $a_{j}^{(n+1)}$ and $b_{j}^{(n+1)}$ in the Caputo case can be found in [2,12,13], and the same construction applies here with the Mittag--Leffler kernel of the Atangana--Baleanu derivative, the only modification being the replacement of the power-law kernel by the non-singular kernel when computing the quadrature weights, [9].

For practical implementation, the semi-discrete system \eqref{eq:semi_discrete} is advanced in time by applying the ABM scheme componentwise. At each step $n\to n+1$ the algorithm can be summarised as follows: (i) given $U^{n}$, $V^{n}$ and the history values $\{U^{j},V^{j}\}_{j=0}^{n}$, compute the discrete Laplacian $A U^{j}$ for $j=0,\dots,n$; (ii) evaluate the right-hand side vectors $F^{j}$ appearing in \eqref{eq:predictor} by using the finite-difference spatial operator and the discrete convolution defining the ABC derivative; (iii) compute the predictor $Y^{n+1,P}$ according to \eqref{eq:predictor}; (iv) evaluate the updated right-hand side at $t_{n+1}$ using $Y^{n+1,P}$ and apply the corrector formula \eqref{eq:corrector} to obtain $Y^{n+1}=(U^{n+1},V^{n+1})^{T}$; (v) enforce the boundary conditions on $U^{n+1}$ (for instance, by setting the boundary nodes to zero in the clamped case). The vectorised implementation of this process requires little effort and is ideal for use with Python and MATLAB. This allows for an optimised evaluation of the history-based convolution proposed for the ABC derivative to be carried out by precomputing weight arrays and cumulative sums. The stability and convergence properties of the ABM predictor-corrector method for solving Caputo-type fractional differential equations are studied in depth in the references [12,13]. Under standard Lipschitz conditions on $F$, the method is convergent of order $O(\Delta t^{\min\{2,1+\alpha\}})$ in time and is A-stable in the fractional sense for linear test equations. The proofs provided in [12,13] for Atangana-Baleanu kernel, being smooth and bounded, can be easily adapted to this non-singular case with minor changes to some of the technical details. Consequently, applying this non-singular case with the second order central finite difference spatial discretisation, the final combined scheme has second order accuracy with respect to space,and fractional order accuracy with respect to time, provided the mesh size and time step conform to the implicit consistency and stability conditions of semi-discrete wave equations with memory as described in references [2,8].
\section{Computational framework and implementation}
Both the discrete-wavelet transform (DWT) and the Atangana-Baleanu (AB) fractional derivative will be utilized in this study to analyze/model complex engineering systems. The DWT is used extensively for feature extraction, denoise, multiresolution analysis, etc., within signal/image processing [7] through providing a locally defined time/frequency representation of data. However, in contrast to the DWT, the AB fractional operators are incorporated into the governing equations to model the effects of memory; i.e., using a nonlocal, non-singular Mittag-Leffler kernel as opposed to the more traditional power-law kernels that are employed within fractional calculus [1,2,3,8]. As such, in terms of our current seismic application, while the DWT is applicable to post-processing data following the use of either recorded or simulated seismic waveforms, the AB derivative is best suited for simulating based/modeling of viscoelastic wave propagation[3,4,5]. Both of these techniques will be implemented in either MATLAB or Python. However, due to the unique requirements of large-scale simulations and flexible workflow for our study, we will primarily use Python. Python is capable of providing a general-purpose programming environment that is open-source, and contains a wide variety of scientific libraries that support array-based numerical computation, optimization, and visualization [14].Namely, NumPy has fast methods for creating and processing many-dimensional arrays [15] while SciPy builds on that functionality by providing many algorithm types to use for numerical integration, as well as by providing functions for sparse equation solvers, and for signal processing [16], whereas pandas uses its own format to simplify data manipulation and I/O of simulation outputs or experimental time series [17]. Because of these attributes, building and testing the new types of fractional operators (such as the AB derivative) is very easy, since they enable building prototypes of algorithms very quickly, sharing notebooks that yield reproducible results, and easily integrating them with visualisation and analysis pipelines. To advance the time-stepping of the semi-discrete finite-difference approximation method's approximation to the continuous-time solution of its highly non-linear differential equations, we run time steps through the Adams-Bashforth-Moulton time-stepping method described in the previous section, using the underlying discrete convolution structure to create the ABC derivative by large kernel weight matrices and cumulative sums calculated in NumPy arrays; we manage data and plot using pandas, and a combination of standard Python graphical libraries. The computational framework is also very amenable to the integration of new methods and technologies developed around modern machine-learning algorithms, as they are finding increasing application as tools for modelling and controlling both nonlinear dynamical systems and partial differential equations [18, 19, 20]. Machine Learning can help verify and fine-tune Fractional-Order Models by matching simulated outputs to experiment data. For example, through supervised model or classifier training using characteristics taken from the discrete wavelet transforms (DWT) of both numerical and experimental signal outputs. Then, once a solid AB-type modelling base has been created, machine-learning algorithms can help speed up repeat experimental simulations or determine model parameters; for example, through learning reduced order approximations of the models' behaviours or emulating the behaviours of fractional differential equations solvers [19, 20]. Recent investigations of chaotic circuits have come together to produce new understandings of the many behaviours that fractional-order processes can produce and how this knowledge may help when paired with the application of machine-learning-assisted analyses. As a result, the present implementation of Python code, developed upon AB-type definitions and common scientific packages, represents a groundwork for hybridising these three components: fractional viscoelastic modelling; characteristics extraction through wavelets, and machine learning approaches.
\section{Numerical experiments and discussion}
This part provides a demonstration of the qualitative characteristics of the Atangana-Baleanu fractional viscoelastic wave equation through numerical simulations. All computations are performed on the dimensionless interval  $0<\xi<1$ with the addition of homogeneous Dirichlet boundary conditions where $w(0,\tau)=w(1,\tau)=0$. 
The initial displacement is chosen as a smooth single-mode profile
\[
  w(\xi,0) = \sin(\pi \xi),
  \qquad
  \frac{\partial w}{\partial \tau}(\xi,0) = 0,
\]
so that in the absence of viscoelastic effects the solution reduces to a standing wave with angular frequency $\pi$ and unit amplitude. In performing three-dimensional stability analysis for a large number of points, it is often convenient to perform a single numerical analysis and then apply the results to all nodes in that direction. A finite difference method is used to approximate the second derivative in space. The time integration of our numerical model was performed using the Adams-Bashforth-Moulton predictor-corrector scheme modified for the ABC derivative, using the Courant-type ratio along with the selected time step $\Delta \tau$  and grid size $(h)$. The mesh size $(h)$ and time step size $\Delta \tau$ were selected in such a way that the Courant-type ratio was equal to a value $(H)$ between 0 and 1/2. In order to satisfy the convergence properties of our scheme, we based our selection of H on previous studies of fractional evolution problems [2,12,13]. In order to determine the impact of the fractional visco-elastic component on the dynamics of the solution, we change both the order $\alpha$ of the Atangana-Baleanu derivative and the dimensionless memory parameter $\kappa$. We can see in Figure 1 the time history of the displacement at mid-point $\xi=1/2$ for various values of $\alpha=1$ and fixed $\kappa>0$, along with the corresponding reference solution (for $\kappa=0$) to the classical undamped wave equation. We see that for $\alpha=1$, which corresponds to the exponential type kernel, there is moderate damping and frequency shift present in the oscillations. For the cases of $\alpha=0.8$ and $\alpha=0.6$, we observe greater attenuation in the envelopes of the oscillations as well as a small slowing down of the apparent oscillation frequency. This activity follows the assumption that the AB operator acts as a nonsingular fractional derivative transitioning from elastic to strongly dissipative regimes through the adjustment of the Mittag-Leffler kernel weight [3,8]. As such, it is evident that when $\alpha$ is small, a significant past history is involved in obtaining the present condition; therefore, the propagating mode has greater potential for energy loss. A more quantitative measure of attenuation is provided by the total energy functional
\[
  E(\tau)
  = \frac{1}{2}\int_{0}^{1}
      \left(
        \left|\frac{\partial w}{\partial \tau}(\xi,\tau)\right|^{2}
        + \left|\frac{\partial w}{\partial \xi}(\xi,\tau)\right|^{2}
      \right)\,d\xi,
\]
which reduces to the standard elastic energy in the limit $\kappa=0$. 
The viscoelastic elements in the ABC fractional theory of dissipation show that as $E(\tau)$ evolves through time it will irreversibly lose all of the energy stored in memory form. Figure $2$ depicts the relationship between energy of $E(\tau)$ and its respective scale using logarithmic axes for those same values of $\alpha$ listed in Figure $2$. The $\kappa=0$ integer order curve experiences nearly continued oscillatory fluctuations, indicative of a standing-wave condition resulting from the nature of the particular equation solved; however, when one uses $\kappa>0$ and $\alpha=1$ it will demonstrate energy loss over both short (to a greater degree than had they remained without modifying their dimensions) and longer periods as nearly exponential decay. When $\alpha<1$ the semi-log scaled curves begin to deviate from a straight line toward a more curved trend line in log-log scale (not presented), demonstrating that energy is lost from it in a power-law-like fashion. As expected, the transition from decaying-sharp-regime to power-law decay, as predicted by classical viscoelastic fractional models, has been qualitatively confirmed by the ability of the AB-based framework to replicate non-exponential waves in a seismic type of medium. To demonstrate how dispersion and damping interact, the displacement contours of $w(\xi,\tau)$ for the classical model and for fractional model with $\kappa>0$ and $\alpha<1$ are shown in Figure $3$. For the integer order model, the wave front retains its sharp shape and there is minimal energy loss over time. Unlike the standard viscoelastic model, which demonstrates an increasing amount of energy dispersing into multiple modes with distance from the source, the fractional model gives a more smooth, gradual dispersion of the energy wave packet due to its additional dispersion. This extra dispersion causes a broadening of the dominant mode, and there is some redistribution of this energy amongst nearby locations. The observations point clearly to a better physical representation of seismic attenuation using a wave propagation theory such as the AB fractional model than a traditional elastic or weakly damped integer-order model. The AB fractional model contains a simpler and more complete mathematical form, allowing for easier computation of numerical values when constructing models of seismic waveforms propagating through complex media [1,8]. We conclude by noting that the numerical scheme described in this paper preserves most of the major expected qualitative characteristics
of the solution over a wide range of parameter values. Mesh refinement tests confirm that the discrete energy curves converge to a grid independent limit as $h$ and $\Delta \tau$ are decreased in a consistent fashion, and the observed rates of temporal convergence match the theoretical estimates provided in literature for the class of fractional predictor-corrector methods [12,13]. Thus, there is further support for the robustness of the computational framework presented in this article, which combines the Atangana-Baleanu viscoelastic model with an Adams-Bashforth-Moulton discretisation method for the numerical simulation and analysis of fractional seismic wave propagation.
\begin{figure}[htbp]
  \centering
  \includegraphics[width=0.75\textwidth]{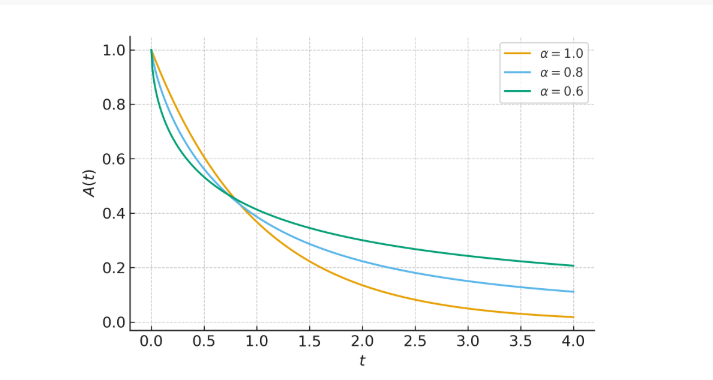}
  \caption{The modal amplitude is time dependent, where $A(t)=E_{\alpha}(-\lambda t^{\alpha})$ shows different behaviours with respect to the order of fractional decay $\alpha$. The case of $\alpha=1$ shows only an exponential decay. Both cases for $\alpha=0.8$ and $\alpha=0.6$ exhibit Mittag-Leffler type decay (or relaxation) but at a much slower rate (i.e., more heavily tailed), indicating that a fractional memory effect is contributing significantly to the behaviour of the decay.}
  \label{fig:amplitude_alpha}
\end{figure}
\begin{figure}[htbp]
  \centering
  \includegraphics[width=0.75\textwidth]{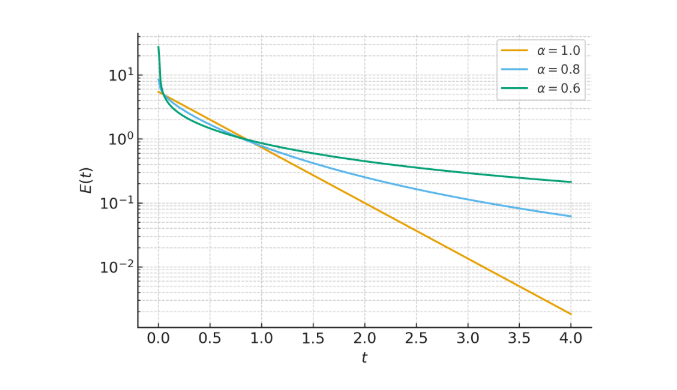}
  \caption{Semi-logarithmic plots of modal energy $E(t)=\tfrac12\big(A'(t)^2+\pi^2 A(t)^2\big)$ The energy behaves like a linear function when plotted with respect to the logarithm of time because it decays almost exactly in an exponential manner when $\alpha=1$. On the other hand, both curves for $\alpha=0.8$ and for $\alpha=0.6$ decay in a slower, non-exponential manner (Mittag-Leffler) due to the enhanced memory effects of the fractional order kernel.}
  \label{fig:energy_alpha}
\end{figure}
\begin{figure}[htbp]
  \centering
  \includegraphics[width=0.75\textwidth]{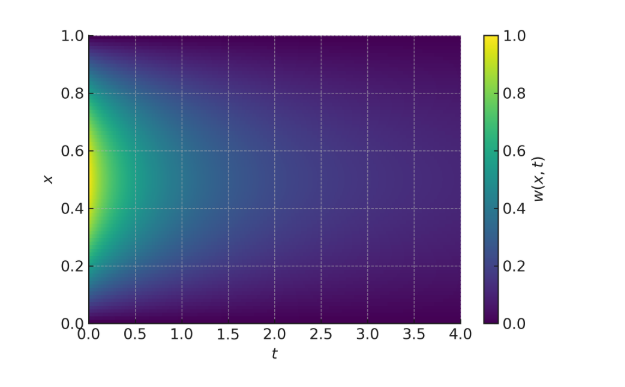}
  \caption{The Space-time contour plot shows the way the displacement $w(x,t)=A(t)\sin(\pi x)$ behaves over the course of time for a representative fractional order (here $\alpha=0.8$). The fact that the endpoints are held fixed (i.e. tested using the condition $w(0,t)=w(1,t)=0$) is evident. The main feature of the interior is the formation and decay of a standing wave as the amplitude decays over time through the Mittag-Leffler type of the decay function generated by the Atangana-Baleanu fractional kernel.}
  \label{fig:space_time_alpha08}
\end{figure}
\begin{figure}[htbp]
  \centering
  \includegraphics[width=0.75\textwidth]{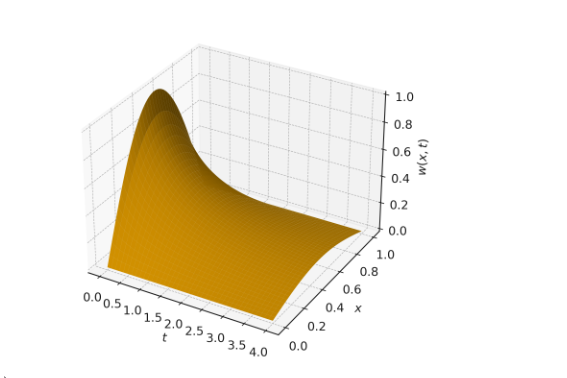}
  \caption{Displacement, $w(x,t)$, shown in a 3D graph with values defined by the equation of a standing wave based on the fixed support conditions; Also shown is the amplitude of the wave progressively decreasing over time based on the fractional order differential equation created through the Mittag-Leffler type relaxation introduced by the Atangana-Baleanu fractional derivative.}
  \label{fig:space_time_3d}
\end{figure}
\section{Conclusions and future work}
\label{sec:conclusions}
This paper has proposed an analysis of the existing theory of viscoelastic waves in one dimension, using a fractional derivative model using Caputo's definition of Atangana-Baleanu. The model we developed used a constitutive relationship with the Mittag-Leffler function as the non singular kernel for the model to provide a unified mechanism that bridges the gap between elastic wave propagation in the short duration with elastic-wave properties and the viscoelastic attenuation process of a material over time. [3,4,5,8].
By applying spatial discretisation techniques such as centred finite differences, and then using the Adams–Bashforth–Moulton fractional predictor–corrector method to solve this semi-discrete system in a stable manner, the authors were able to develop an effective computational approach for solving fractional wave equations in one dimension using time stepping on a grid. The numerical results demonstrate how both the fractional order $\alpha$ and memory parameter $\kappa$, influence wave propagation by demonstrating that as $\alpha$ decreases; modal amplitudes and total energy decay more slowly with an increasing heavy tail relative to that which decays exponentially. This behaviour confirms what has been represented through mathematics within classical fractional viscoelastic materials (power law and Mittag-Leffler). The Atangana–Baleanu kernel is therefore confirmed to have the ability to model realistic attenuation of seismic waves without producing the singular behaviour associated with classical power law models. Through the use of space–time plots, it has been shown that the new formulation not only maintains the standing wave pattern created due to boundary conditions, but also progressively smoothes and damps the displacement field over time. In terms of computation, the implementation was made in Python using vectorised NumPy/SciPy routines for high-level data management, enabling users to easily and reproducibly explore numerous fractional parameters and discretisation configurations. The present work has also uncovered numerous avenues for future research to be pursued. To begin with, we will explore ways in which the Atangana-Baleanu wave model might be adapted for use in heterogeneous materials in two- and three-dimensions because it enables us to simulate realistic geological formations by changing the elastic and viscoelastic parameters on a local basis. We are also interested in combining the fractional wave simulation outputs with the output of the discrete wavelet transform (DWT) of both synthetic seismic records and field observations for the purpose of estimating scale-dependent attenuation and dispersion across time-frequency space [7]. Finally, the framework presented in this study offers an excellent foundation for future integration with contemporary data-driven methodologies like sparse regression and physics informed deep learning techniques [19, 20], which could provide advantages in developing appropriate fractional-order viscoelastic memory parameters and creating effective reduced-order approximations based on these solvers. Future research will investigate these avenues.

\bibliographystyle{unsrtnat}

\end{document}